\documentclass{optica-article}

\journal{opticajournal} 

\articletype{Research Article}

\usepackage{lineno}
\usepackage{rotating}  

\usepackage{graphicx}

\usepackage{setspace}

\usepackage{amsmath}
\usepackage{float}   

\begin{document}

\title{Spectral DiffuserScope: a compact snapshot hyperspectral microscope}

\author{Neerja Aggarwal\authormark{1*}, Eric Markley\authormark{2*}, Kyung Chul Lee\authormark{3}, Seung Ah Lee\authormark{3,4}, Junghyun Bae\authormark{5}, Nakkyu Baek\authormark{3}, Wook Park\authormark{5}, Min Sung Cho\authormark{5}, Youngbin Lim\authormark{5}, Polly Fordyce\authormark{6}, Stephanie Eberly\authormark{7}, Lydia L. Sohn\authormark{7},  William D. Houck\authormark{8}, Kristina Monakhova\authormark{9},  Laura Waller\authormark{1+}}
\address{
\authormark{*} authors contributed equally\\
\authormark{1}Department of Electrical Engineering and Computer Sciences, University of California, Berkeley, USA \\
\authormark{2}UCB/UCSF Joint Graduate Program in Bioengineering, University of California, Berkeley, USA \\
\authormark{3} School of Mechanical and Aerospace Engineering/SNU-IAMD, Seoul National University, Seoul, Republic of Korea
\authormark{4} Department of Mechanical Engineering, Seoul National University, Seoul, Republic of Korea
\authormark{5} Department of Electronic Engineering, Kyung Hee University, Yongin-si, South Korea 

\authormark{6}Department of Bioengineering and Genetics, Stanford University, Stanford, USA \\
\authormark{7}Department of Mechanical Engineering, University of California, Berkeley, USA \\
\authormark{8} VIAVI Solutions Inc, Santa Rosa, USA \\
\authormark{9}Department of Computer Science, Cornell University, Ithaca, USA \\
\authormark{+}correspondence: neerja@alum.mit.edu, waller@berkeley.edu
} 


\begin{abstract*} 
Hyperspectral fluorescence microscopy enables important biological and clinical applications, but conventional systems are bulky or require scanning, limiting temporal resolution and throughput. We introduce a computational snapshot hyperspectral microscope that uses compressed sensing to achieve higher spatial-spectral resolution than traditional snapshot systems. Our device is compact ($\sim$ 15 cm $\times$ 6 cm $\times$ 6 cm) and easily attaches to standard fluorescence microscopes. We benchmark our system against existing snapshot methods through simulations to evaluate its spatial and spectral performance. Experimental imaging of fluorescent beads, labeled cells, and lanthanide hydrogel beads demonstrates a practical, high-throughput solution for hyperspectral microscopy in biological and clinical applications.
\end{abstract*}

\section{Introduction}
Hyperspectral microscopy aims to capture a high-resolution spectrum for each spatial location in an image.  Having both spatial and spectral information about a sample is valuable in many applications in fluorescence microscopy, such as cellular imaging. For example, hyperspectral microscopy enables linear unmixing of dozens of fluorophores, allowing denser labeling of cell structures~\cite{McNamara2024-vp}, which is useful for genomic phenotype screens~\cite{Sivanandan2023-lq}.  Traditionally, multichannel fluorescence imaging (with only a few spectral channels) is used for high-throughput, quantitative cellular imaging~\cite{Zanella2010-gw}, but expanding to a hyperspectral system has the potential to produce more useful, richer datasets~\cite{timlin_hyperspectral_2004}. Additionally, hyperspectral microscopy enables bead-based bioassays in which small fluorescent beads are used to detect and quantify small molecules such as nucleic acids and proteins.  Hyperspectral microscopy allows simultaneous detection and differentiation of multiple bead types based on their spectral signatures, enabling highly multiplexed assays~\cite{Yuan2018-xr, Feng2020-rx, Nguyen2017-pu, Gerver2012-ye}. In the work by Feng \textit{et al.}, a basis of six different spectra was used to create ratiometric lanthanide hydrogel beads that required six emission channels for identification.  Hyperspectral microscopy can enable the identification of even more codes, further increasing throughput.   

Most hyperspectral imaging systems collect a 2D spatial + 1D spectral datacube by scanning through space or spectra.  For example, in a pushbroom system, a single spatial line in the object is spectrally dispersed using a grating onto a 2D sensor.  The object is slowly scanned through the other lateral dimension to generate a 3D spatio-spectral datacube~\cite{arablouei_fast_2016, song_hyperspectral_2024}. Alternatively a tunable spectral filter can be placed before the sensor to scan through wavelength channels instead of spatial locations~\cite{favreau_tunable_2013}.  In both cases, multiple measurements must be acquired to generate the 3D hyperspectral datacube, requiring long acquisition times per sample. These extended acquisition times fundamentally limit throughput in high-content screening applications and prevent the collection of large-scale datasets needed for modern machine learning approaches to biological image analysis. Fast, snapshot hyperspectral acquisition is therefore critical for enabling data-driven discovery in biological systems.
Snapshot hyperspectral imagers can capture the whole spatial-spectral datacube in a single exposure, potentially speeding up acquisition time and enabling applications with dynamic samples. Several research systems have demonstrated snapshot fluorescence microscopy, as summarized in Supplementary Table~\ref{tab:hyperspectral}.  These systems use either filters which allow customizable channel selection or dispersive prisms which improve light throughput to separate out the wavelength channels. For example, the systems in References ~\cite{Lavagnino2016-us}, ~\cite{Orth2014-fx}, and ~\cite{Zhang2015-lp} all use prisms to disperse the light from the scene onto a 2D sensor relying on scene sparsity to capture the entire datacube. This approach allows for live cell imaging, however, it relies on large setups with limited spectral channels. Wu \textit{et al.} uses a multi-camera array with a different color filter on each camera to obtain volumetric multispectral imaging~\cite{Wu2016-if}, but this strategy does not scale well to high-resolution spectral data. Our design uses a filter array for a more compact and accessible approach. Like our approach, previous work has also used compressed sensing to recover a 3D datacube from a 2D measurement. For example, the coded aperture snapshot spectral imaging (CASSI) system captures the full spectral datacube by using a prism to disperse the different wavelengths and an aperture to code the information onto a 2D sensor~\cite{Cull2010-ky}. Inspired by this, we present a different compressed sensing approach using a diffuser for multiplexing. 

In this paper we present a new compact, computational hyperspectral microscopy system that enables snapshot imaging with standard benchtop epifluorescence microscopes. Building on our prior work in hyperspectral photography~\cite{Monakhova2020-gd}, the system consists of just three components that attach to a microscope’s output port. It reconstructs a hyperspectral datacube with more voxels than sensor pixels using compressed sensing, achieving ~3$\times$ higher spatial resolution than traditional filter array-based snapshot systems. The current design captures 64 spectral bands ($\sim$8 nm resolution) across the visible to near-infrared and in future work these could be customized for specific applications. We release the hardware and reconstruction software as open source to support adoption and extension. Fluorescent beads, commonly used in bioassays, demonstrate accurate spatial and spectral recovery. We also discuss key limitations in signal strength and reconstruction quality, and propose paths forward.

\section{Methods}
Our imaging architecture employs multiplexing and compressed sensing to reconstruct a hyperspectral data cube from a single sensor measurement with relatively simple hardware. Compressive sensing encodes a high-dimensional scene into the 2D measurement intelligently, enabling accurate recovery of the full datacube with significantly fewer samples than traditional approaches. To achieve this, our system leverages custom diffuser with a large point spread function (PSF) with sharp features, mapping each point in the sample to multiple points on the sensor. This design ensures complete spectral sampling at every spatial location while preserving high-frequency details, allowing for high spatial and spectral resolution reconstructions.

\begin{figure}[htb]
    \centering
    \includegraphics[width=1\linewidth]{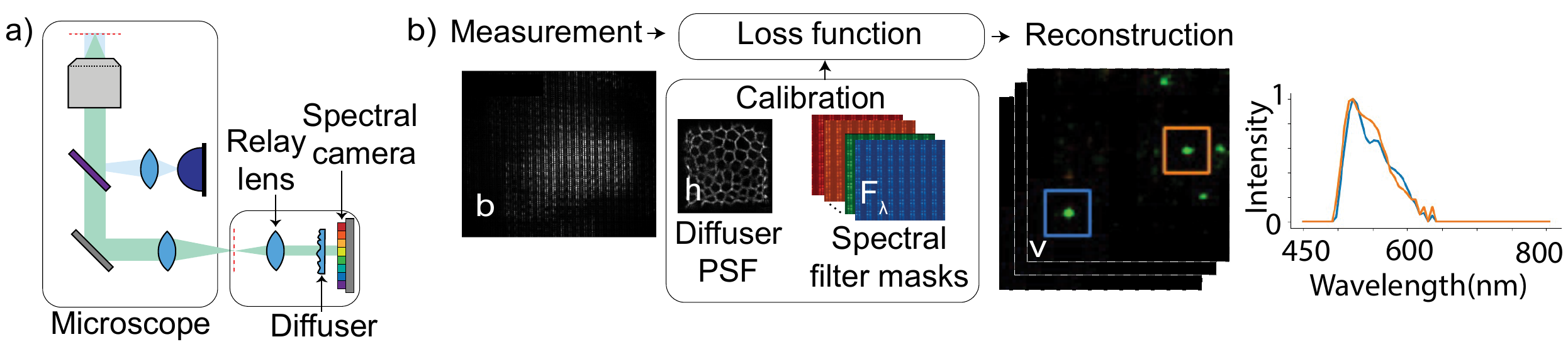}
    \caption{a) Our snapshot hyperspectral microscope consists of a relay lens, diffuser and spectral sensor placed at the output port of a commercial fluorescence microscope. The relay lens images the Fourier (pupil) plane of the objective onto the diffuser to shape the system's point spread function (PSF). The spectral camera has a 64-channel spectral filter array placed on top of a monochrome sensor. b) To reconstruct hyperspectral images from a single captured measurement, our algorithm starts with a monochrome 2D sensor measurement and passes it through an optimization problem with calibration data to recover the hyperspectral datacube.  The recovered spectral scene here is fixed fluorescently-labeled \textit{in vitro} cells with the spectra plotted to the right. }
    \label{fig:overview}
\end{figure}

\begin{figure}[tbh]
    \centering
    \includegraphics[width=1\linewidth]{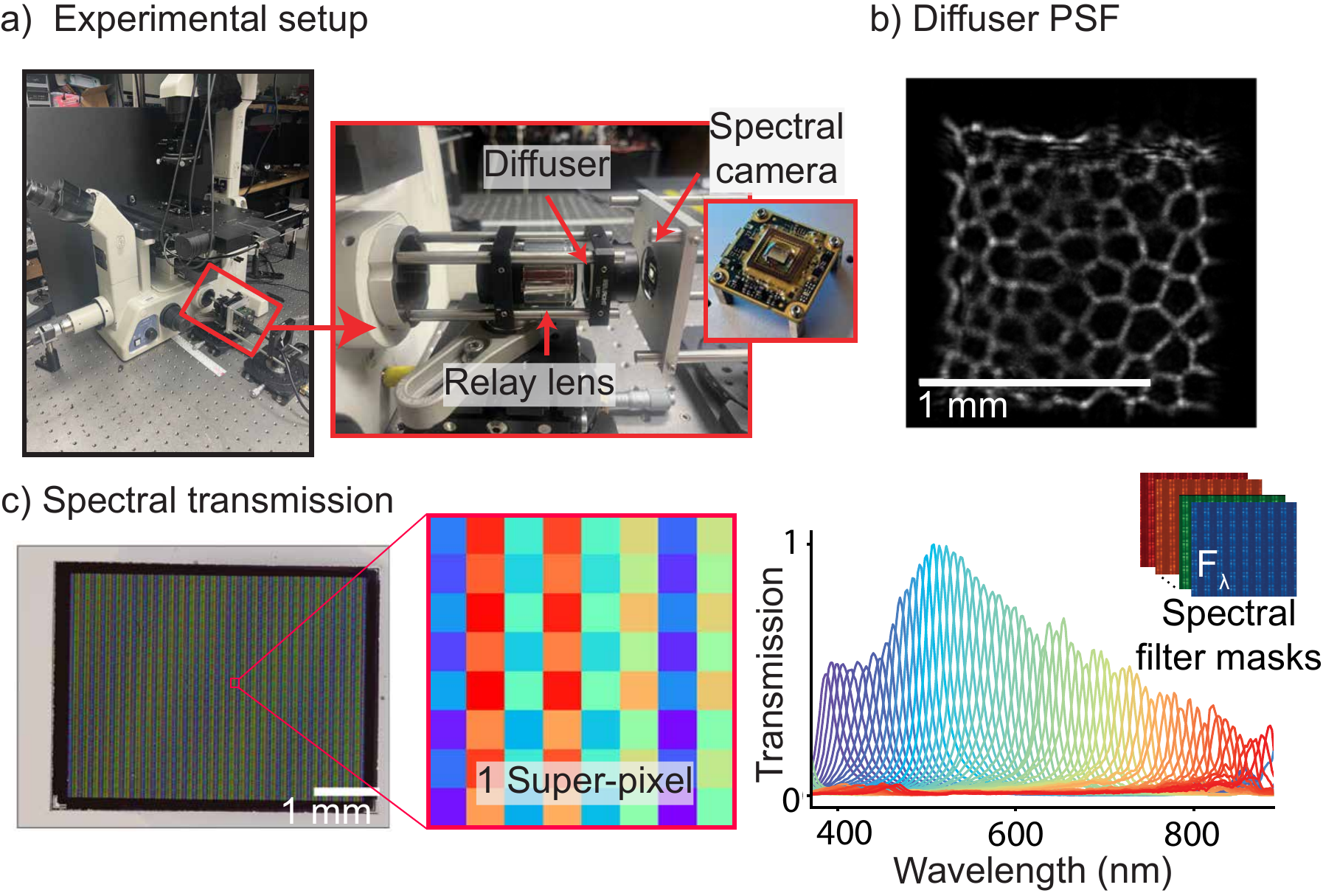}
    \caption{(a) Experimental setup with our hyperspectral camera mounted on the side port of a Nikon TE300 microscope, consisting of a relay lens, diffuser, and spectral camera. The spectral camera is a board-level CMOS sensor with the cover glass removed and a spectral filter adhered to its surface. 
    (b) The diffuser's PSF, captured from a 5 µm fluorescent bead sample, is shown on the unfiltered pixels of the sensor. 
    (c) The spectral transmission matrix was obtained by scanning the source wavelength with a monochromator. The false color representation of the transmission matrix is shown along with a zoomed image displaying the 64 filters within a single super-pixel. To the right, the measured transmission spectrum of each filter is shown.}
    \label{fig:calibration}
\end{figure}

The compact optical setup (Fig.~\ref{fig:overview}) integrates easily with a standard benchtop fluorescence microscope. Emission light from the sample is directed through the microscope’s objective and into the hyperspectral imager, which attaches to the sideport. Measuring approximately 15 cm $\times$ 5 cm $\times$ 5 cm, the imager is easily aligned using a standard sideport adapter (Fig.~\ref{fig:calibration}a). This architecture builds upon our prior work with the Spectral DiffuserCam~\cite{Monakhova2020-gd}, incorporating a relay lens improves signal throughput by closely matching the size of the Fourier plane to the diffuser aperture and simplifying integration with existing fluorescent microscopes. 

The imager comprises of three main components: a relay lens, a custom-engineered diffuser, and a spectral camera. The relay lens transfers the Fourier plane of the microscope’s objective to an accessible location at the output port.  The custom-engineered diffuser is positioned at the relayed Fourier plane to maintain shift invariance. The diffuser shapes the system’s PSF, ensuring it spans multiple super-pixels on the spectral camera to achieve full spectral sampling. The PSF’s sharp caustic features encode the lateral position of each point source while preserving high-frequency information. Thus, the diffuser allows the system to capture both spatial and spectral details simultaneously from the sample.  The hyperspectral camera is positioned at the focal plane of the diffuser; it includes an off-the-shelf CMOS sensor with a 64-channel spectral filter array adhered directly to the sensor plane. The filter array consists of 8$\times$8 super-pixels, each containing 64 unique spectral filters, with each super-pixel spanning dozens of sensor pixels.

\subsection{Forward Model}

To computationally reconstruct the sample's hyperspectral datacube, it is necessary to first model the measurement formation of the imaging system. The system captures fluorescence emission, which is inherently incoherent, ensuring that light emitted from different points in the sample does not interfere. This results in a measurement model that is linear in intensity, where the total signal at the sensor is a weighted sum of contributions from all points in the scene. Because the diffuser is in the Fourier plane, the system will further be shift-invariant, enabling the use of a convolution-based forward model similar to the derivation in our previous work~\cite{Monakhova2020-gd}. We model the captured sensor measurement, ${\bf b}[x,y]$, as:

\begin{equation}
\label{eq:fwmodel}
{\bf b} =
\sum\limits_{\lambda = 0}^{K - 1} {{\bf F}_\lambda}[x,y] \cdot
\text{crop}({\bf h}[x,y]\mathop *\limits^{[x,y]} {\bf
v}[x,y,\lambda]),
\end{equation}

\noindent where the sample’s hyperspectral datacube is denoted as ${\bf v}[x,y,\lambda]$, $x$ and $y$ are the spatial dimensions and $\lambda$ represents the spectral dimension. The diffuser’s PSF, ${\bf h}[x,y]$, is convolved with the sample ${\bf v}[x,y,\lambda]$ over the spatial dimensions, $x$ and $y$. In our setup, the PSF does not vary significantly with wavelength, allowing ${\bf h}[x,y]$ to be approximated as independent of $\lambda$. The resulting convolved image is further filtered by the spectral filter array which controls the transmission of each wavelength channel.  This is modeled by element-wise multiplication with ${\bf F}[x,y,\lambda]$, the calibrated spectral transmission matrix. Since the individual wavelength channels are incoherent, their intensity contributions can be summed across $K$ spectral channels to compute the estimated measurement from an estimate of the hyperspectral datacube.

The forward model for microscopy has an extra step beyond that of photography in that the magnification  of the microscope must be taken into account when sizing the reconstructed pixels.  The effective magnification of our system combines the magnification of the microscope objective with the demagnification of the hyperspectral imager:

\begin{equation}
\label{eq:meff}
    M_{\text{eff}} = M_{\text{obj}} \cdot \frac{f_{\text{diffuser}}}{f_{\text{relay}}}.
\end{equation}

The ratio of focal lengths between the relay lens and the diffuser determines the demagnification of the image. This demagnification is critical to fit the entire field of view onto the spectral filter array (~4~mm).  With our chosen values of $ f_{diffuser} = 9$ mm and $f_{relay} = 39 $mm, the effective magnification is 0.9$\times$ when using a 4 $\times$ objective.  The effective magnification increases to 2.3$\times$ for a 10$\times$ objective. This effective magnification along with the sensor pixel size is used to size the reconstructed object pixels: 

\begin{equation}
\label{eq:pixsize}
    \Delta x_{recon} = \Delta x_{sensor}/M_{eff}.
\end{equation}

\subsection{Reconstruction Algorithm} \label{sec:recon_alg}

Since our inverse problem represents an under-determined linear system with infinitely many possible solutions, we constrain the solution space using multiple complementary priors. We apply native sparsity priors to exploit the inherent spatial sparsity typical of fluorescent microscopy scenes. Additionally, we enforce piecewise smoothness through penalties on first spatial derivatives in the lateral dimensions, while penalties on second spectral derivatives encourage smoothly varying spectra. We further enforce a limited dictionary of spectral components to make up the overall datacube.  This is relevant for fluorescence imaging where only a known number of different dyes are imaged within a scene.  These constraints, rooted in compressed sensing theory, allow accurate reconstruction by leveraging the structure and properties of the hyperspectral datacube, even under the limited measurements provided by the system.  

The hyperspectral datacube, ${\bf v}[x, y, \lambda]$, is first modeled separably as a decomposition of the spatial and spectral components:

\begin{equation}
{\bf v}[x, y, \lambda] = \sum_{k=1}^{N_f} {\bf V}_k[x, y] \cdot {\bf U}_k[\lambda],
\end{equation}

\noindent where ${\bf V}_k[x, y]$ represents the spatial weight map for the $k$-th fluorophore, and ${\bf U}_k[\lambda]$ represents its spectral profile, capturing its emission spectrum. An additional $U_k$ component is included to account for any additional signal such as background or illumination leakage. Consequently, ${N_f}$ equals the number of fluorophore types plus one.

The spatial weights ${\bf V}_k[x, y]$ enable the model to gradually learn which spectral component belongs at each spatial location, under the assumption that only one fluorophore type is present at each pixel. Rather than making hard binary assignments from the start, we use a softmax operation with temperature $\tau$ to create soft assignments from the raw weights ${\bf W}[x, y, k]$. This approach allows smooth optimization toward the ultimate goal of one dominant component per location, while the reweighting term ${\bf \alpha}[x, y, k]$ controls intensity scaling.  This process is given by the equation:

\begin{equation}
{\bf V}_k[x, y] = {\bf \alpha}[x, y, k] \cdot \text{softmax}\left(\frac{{\bf W}[x, y, k]}{\tau}\right).
\end{equation}

Initially, the temperature parameter $\tau$ allows multiple spectral profiles to contribute to a single spatial location, enabling flexibility in the early stages of reconstruction. As optimization progresses, $\tau$ is gradually decreased, sharpening the softmax distribution and enforcing the selection of a single dominant spectral profile per spatial location, consistent with the assumption that only one fluorophore is present at each pixel.  We deem this our low-rank, one-hot reconstruction: a decomposition that represents the high-dimensional spectral data using a small number of basis spectra (low-rank), with each spatial location assigned to exactly one spectral component (one-hot), to reconstruct the original measurements.

The reconstruction algorithm optimizes ${\bf W}[x, y, k]$, ${\bf \alpha}[x, y, k]$, and ${\bf U}_k[\lambda]$ to refine the hyperspectral datacube ${\bf v}[x, y, \lambda]$. This process incorporates prior knowledge, such as the known number of fluorophores, $N_f$, and the sparsity of active fluorophores, to guide the reconstruction toward physically meaningful results that align with the observed sensor measurements. Note that $N_f$ serves as an upper limit of how many spectral components the algorithm learns.  We do not make assumptions regarding the specific shape or peak location of the fluorescence spectra, maintaining generalization across samples. 

The reconstruction is formulated as the minimization of a loss function that balances fidelity to the experimental measurement with regularization terms promoting sparsity as well as spatial and spectral smoothness:

\begin{equation}
\mathcal{L} = \| \mathbf{b} - \mathbf{\hat{b}} \|_2^2 + \lambda_{\text{TV}} \, \text{TV}(\mathbf{v}) + \lambda_{\text{spectral}} \| \nabla^2_{\lambda} \mathbf{U} \|_1 + \lambda_{\text{sparsity}} \| \mathbf{\alpha} \|_1 + \lambda_{\text{diversity}} \cdot \| \mathbf{U}^\top \mathbf{U} - \mathbf{I} \|_F^2.
\end{equation}

The first term ensures consistency between the observed measurement $\mathbf{b}$ and the estimated measurement $\mathbf{\hat{b}}$ computed using Eq.~\ref{eq:fwmodel}. The second term promotes spatial smoothness by penalizing high-frequency variations in ${\bf v}[x, y, \lambda]$ across the $x$ and $y$ dimensions. The third term enforces smoothness in the spectral profiles by penalizing large second derivatives along the spectral dimension. The sparsity term, $\lambda_{\text{sparsity}} | \mathbf{\alpha} |_1$, reflects the native sparsity of the scene, enforcing the assumption that most pixels correspond to regions without active fluorophores.  Finally, the last term of the loss function enforces orthogonality between learned spectral components by penalizing off-diagonal elements of their Gram matrix. This constraint reflects the standard practice in fluorescence microscopy of selecting fluorophores with minimal spectral overlap to enable robust spectral unmixing. By encouraging orthogonal basis spectra, we reduce redundancy and improve the interpretability of the learned components. The parameters $\lambda_{\text{TV}}$, $\lambda_{\text{spectral}}$, $\lambda_{\text{sparsity}}$, and $\lambda_{\text{diversity}}$ are hyperparameters that control the relative influence of each prior in the loss function.

By iteratively minimizing this loss function, the algorithm progressively refines the hyperspectral datacube, ensuring an accurate reconstruction that captures both the spatial and spectral properties of the sample. This optimization problem is solved using JAX's autograd functionality with the Adam optimizer from Optax \cite{Kingma2014-sx}, configured with $\beta_1 = 0.9$ and $\beta_2 = 0.999$, as per the standard momentum parameters.

\subsection{Comparison with Alternative Optical Designs}
To demonstrate the advantages of our multiplexing pattern PSF approach, we compare reconstructions of a simulated object across four different optical configurations that illustrate the fundamental tradeoffs in snapshot hyperspectral imaging. Each system uses the same spectral filter array sensor, but employs different optical strategies to address the core challenge: how to achieve both high spatial resolution and complete spectral sampling simultaneously with a single exposure.  In all cases, we simulate fluorescent beads spelling CAL, with the spectral profiles of the lanthanide dyes used in Reference \cite{Feng2020-rx} and \cite{Nguyen2017-pu}.  The fluorescent beads are 2.4 micron in diameter at the object plane. For simulation purposes only, the spectral filter array is pixel aligned to the sensor and has ideal transmission curves with 100\% transmission at the peak wavelength and 1\% at all others.  This is in idealized version of the spectral filter used for experiments.  Each individual spectral filter is 20 $\times$ 20 micron squares arrayed into  8 $\times$ 8 superpixels.  Each superpixel is 160 $\times$ 160 microns. 

\begin{figure}[htb!]
    \centering
    \includegraphics[width=1\linewidth]{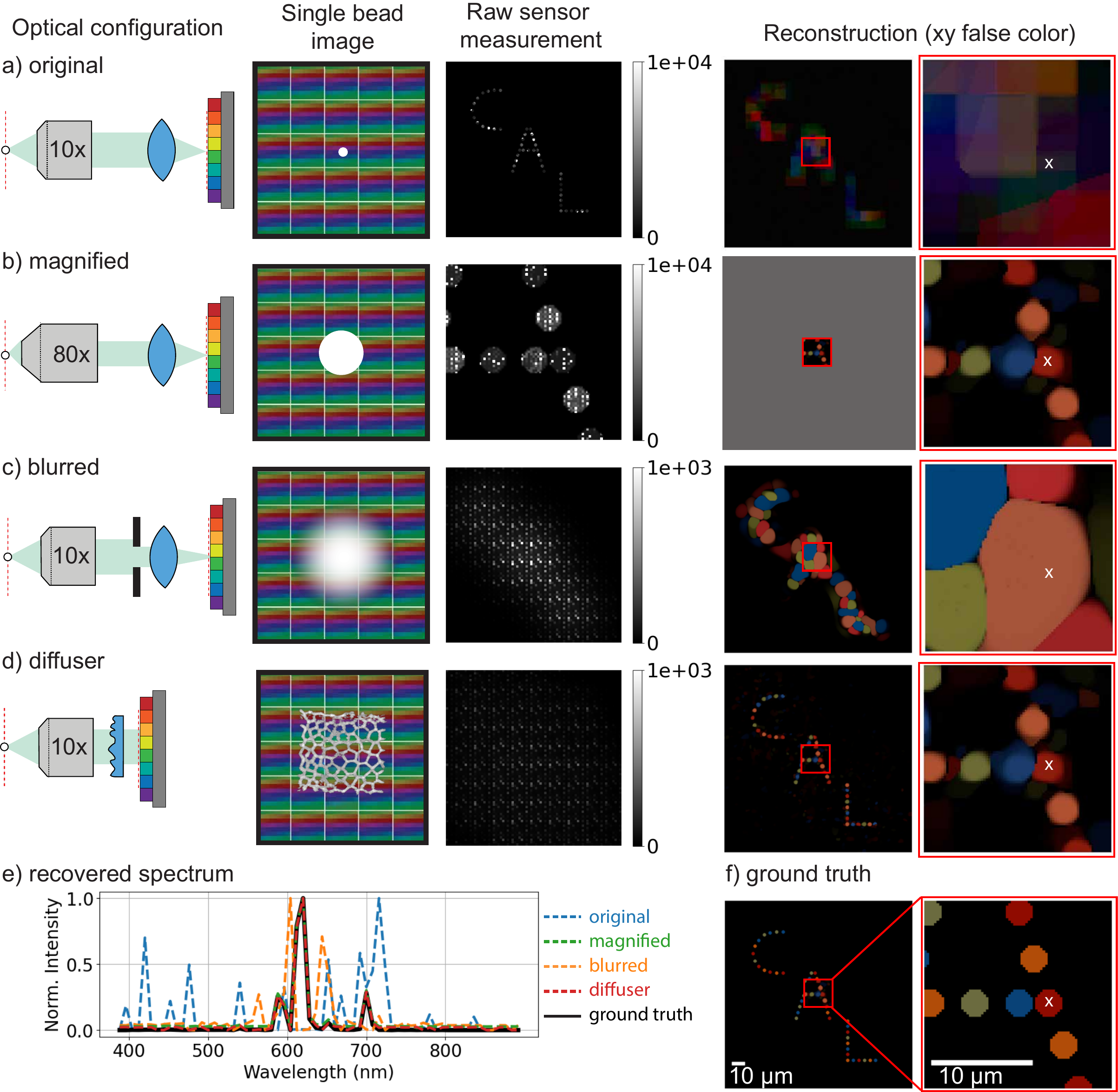}
    \caption{Comparison of optical configurations for snapshot hyperspectral microscopy with a spectral filter array based camera, in simulation.  a) Placing the spectral filter and sensor in the imaging plane of the microscope results in poor spectral recovery, since the light from any given spatial position is not spread out over many spectral channels.  b) A higher magnification objective will spread out a point source's information across many spectral channels, preserving spectral and spatial information, but at the expense of field of view.  c) A purposely blurred PSF can also spread the light across multiple filters but results in poor spatial resolution in the recovered scene.  d) Our approach: a diffuser is placed in the Fourier plane of the microscope to spread the light across multiple filters on the spectral camera, giving good spectral reconstructions without severely sacrificing spatial resolution, since the PSF maintains high spatial-frequency content.  e)  Recovered spectra from the different configurations shows our diffuser approach captures the best combination of resolution, field-of-view, and spectral information.  f) The ground truth sample used for the simulation - a scene of spectrally-encoded hydrogel beads. Results are shown in false-color contrast stretched for visualization (gamma factor = 0.8) with zoom-ins of the central patch in the far right column. All scale bars = 10 $\mu$m.}
    \label{fig:simulation}
\end{figure}

In the first case, as shown in Fig.~\ref{fig:simulation}a, the spectral camera is placed directly in the image plane of a 10$\times$ microscope system.  The object is imaged onto the sensor, however, the image of each bead only lands on a few of the spectral filters thus has insufficient sampling to recover the full spectrum.  We attempt to use interpolation at each spectral channel to mimic demosaicing algorithms and recover the full spatio-spectral scene.  However, there is insufficient information at each spectral channel and the recovered scene does spatially or spectrally match the ground truth.  

In Fig. \ref{fig:simulation}b, we increase the magnification to 80$\times$ to allow a spectrally homogeneous bead to land on the entire superpixel ensuring full spectral sampling and good spatial resolution, but with a much smaller field of view.  We follow the same procedure as the first case to recover the spatio-spectral scene.  

In Fig. \ref{fig:simulation}c, we return to the lower-magnification 10$\times$ objective and add an aperture in the pupil plane to reduce the system's numerical aperture and make the PSF large enough that each bead's light lands on the entire super-pixel.  We follow the low-rank reconstruction process outlined in Section \ref{sec:recon_alg} using a Gaussian PSF and $N_f = 4$.  The reconstructed image shows poor spatial resolution and inaccurate spectra.  This is because a small shift in the bead position doesn't drastically change the measurement on the sensor when using a Gaussian PSF, hence the inverse problem is poorly conditioned to solve. 

Finally in Fig. \ref{fig:simulation}d, we show PSF shaping using the diffuser placed in the Fourier plane, with the spectral fileter and sensor placed 9~mm after.  The overall system is smaller due to the elimination of the tube lens.  The diffuser's PSF preserves high frequency information in the scene while spreading out each bead's light across sufficient spectral filters to enable full spectral sampling. Because of this, the reconstructed object more closely matches the ground truth shape and spectra, demonstrating high spatial and spectral resolution. 

\section{Experimental System Design}\label{sec:methods:spectral_camera_assembly}
A custom hyperspectral camera was constructed in the lab using a board-level CMOS sensor (The Imaging Source, DMM 37UX178-ML) with a resolution of 3,072$\times$2,048 pixels (6.3 MP) and a pixel pitch of 2.4~µm. To enable integration with the spectral filter array, the sensor's cover glass was removed (Wilco Imaging, Sacramento, CA). The hyperspectral filter array was sourced from Viavi Solutions (Santa Rosa, CA) and consists of an 8$\times$8 grid of individual Fabry–Perot filters (10~µm $\times$ 10~µm each), fabricated as a dielectric stack of optical coatings on a glass substrate.

To assemble the spectral camera, the spectral filter array was bonded to the CMOS sensor using optical adhesive (Norland 61). A small drop of adhesive was applied to the sensor surface, and the filter array was carefully lowered, with its optical coating side facing the sensor. The placement ensured proper optical interfacing. The adhesive was cured under an ultraviolet lamp for 15 minutes to secure the bond.

The optical set up consisted of 3 components (a relay lens, a custom diffuser, and spectral camera) added onto the side port of a standard fluorescence microscope (Nikon TE300).  The relay lens (Thorlabs LSM03-VIS Scan Lens, EFL = 39~mm) was aligned to create a 2$f$ system between the side port's image plane and the diffuser plane. The spectral camera was placed 9$~mm$ after the diffuser, approximately at its caustic plane ($f_{\text{diffuser}} = 9$~mm).  The optical arrangement is shown in Fig.~\ref{fig:overview}a. The board-level camera was mounted onto a custom aluminum fixture to integrate it into a cage rod assembly, as illustrated in Fig.~\ref{fig:overview}a. The complete parts list and assembly instructions are provided in Supplementary Information \ref{sec:supp_methods}.

\subsection{Diffuser Design}

To multiplex light from the sample, a custom diffuser was designed to spread light across the spectral filters while preserving signal-to-noise ratio (SNR) and resolution. Previous work \cite{Monakhova2020-gd} used an off-the-shelf Luminit 0.5° engineered diffuser, which produced a sharp caustic pattern. However, the light intensity varied along the ridges of the pattern, and significant light was directed into the regions between the bright ridges, decreasing SNR and wasting dynamic range of the sensor.  In fluorescence microscopy, compared to photography, total photon emission from each point in the scene is orders-of-magnitude lower.  Thus, we want to spread the photons out to as few pixels as strictly necessary to achieve a high contrast PSF and limit shot noise, while maintaining sufficient multiplexing to ensure full spectral sampling. 

To address these limitations, the custom diffuser was designed to distribute light more uniformly across the ridges, improving the evenness of light distribution among the spectral filters and enhancing spectral conditioning. Multiple pattern densities and focal lengths were fabricated and tested using the methods outlined in previous work~\cite{Lee2023-os}. The target PSF pattern was generated using a random generation algorithm, with a density matched to the spectral filter size to ensure uniform information spread. The diffuser’s height profile was subsequently optimized using a differentiable design algorithm to achieve isotropic sharpness across the entire PSF pattern at a specific focal plane~\cite{Lee2023-os}.

The final diffuser design had a focal length of 9~mm, with a Voronoi pattern featuring an average seed density of 44~dots/mm$^2$ and a mask size of 1.4~mm~$\times$~1.4~mm \cite{Lee2023-os}. The 9~mm focal length was chosen to balance two competing requirements: shorter focal lengths provide better angular resolution (larger PSF shifts on the sensor for given sample displacements), while longer focal lengths are able to better preserve high-frequency spatial details through sharper point spread functions. Additional details on PSF comparisons and diffuser specifications are provided in Supplementary Information \ref{sec:supp_methods}.
\subsection{Calibration}\label{sec:methods:calibration}

Calibrating the diffuser's PSF and the spectral filter array's transmission is essential for accurate image formation modeling in the reconstruction algorithm. Spectral calibration was performed immediately after assembling the spectral camera and prior to microscope alignment using a Cornerstone 130 monochromator and following the procedure described previously~\cite{Monakhova2020-gd}. The monochromator's output slit was set to produce an 8 nm full-width half-maximum (FWHM) beam, scanned from 350 nm to 900 nm. The captured images were normalized by the monochromator's output power. After coupling the diffuser and camera to the microscope via the relay lens, the diffuser PSF was captured by imaging a 5~µm fluorescent bead placed in the sample plane and imaged onto a region of the sensor not covered by the spectral filter, using a 4$\times$ objective.

\section{Experimental Results}
We start by characterizing our system resolution experimentally. As this is a computational imager with a nonlinear reconstruction algorithm, its resolution is affected by the: 1) sharpness of the diffuser's PSF, 2) focal length ratio between the relay lens and the diffuser, 3) condition number of the forward model matrix, and 4) any priors used in the reconstruction. 

\begin{figure}[htb!]
    \centering
    \includegraphics[width=1\linewidth]{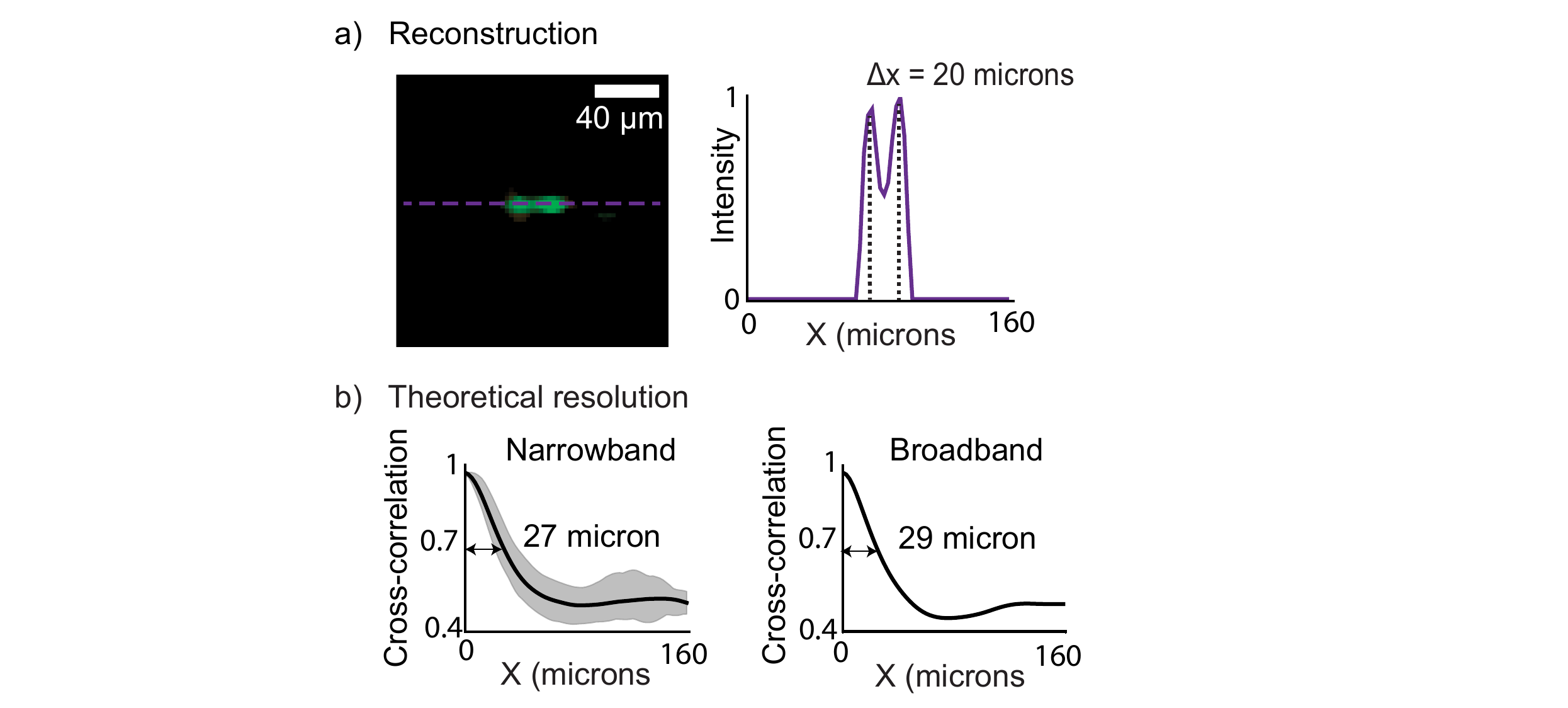}
    \caption{Two-point resolution test demonstrating sub-super-pixel resolution. (a) Experimental reconstruction of a two-source sample, with a spatial cross-section along the dashed purple line demonstrating separation of 20 $\mu$m. (b) Theoretical resolution defined at 70$\%$ peak cross-correlation of simulated measurements from two point sources for both a narrowband source (range and average plotted) and a broadband (uniform spectra) source.}
    \label{fig:two-point}
\end{figure}

Theoretical resolution was calculated as the cross-correlation between simulated measurements of a shifting point source.  The simulated measurements were obtained via the imaging forward model in Eq.~\ref{eq:fwmodel} using experimental calibration data for the diffuser PSF, $h[x,y]$, and the spectral transmission matrix, $F[x,y,\lambda]$.  We simulated measurements from both narrowband and broadband (uniform spectra) point sources.  Figure \ref{fig:two-point}b shows the obtained theoretical resolution defined at 70\% of the peak cross-correlation.  This happens at 0.17 super-pixels (27 microns) on average across wavelengths for narrowband and 0.18 super-pixels (29 micron) for uniform broadband points, similar to ~\cite{Monakhova2020-gd}.  The narrowband theoretical resolution varies slightly with wavelength due to the effective PSF erasure from the spectral transmission matrix.  Joint design of the diffuser PSF and spectral filter may help with achieving consistent resolution across wavelength. 

The system's two-point resolution was also measured experimentally with a 4$\times$ objective. Fluoromax-dyed aqueous green and orange beads (ThermoFisher) were diluted in water and a small droplet was placed on a glass slide, dried, and imaged. Due to the difficulty of precisely placing fluorescent beads onto a glass slide, a single 10 $\mu$m bead was imaged, shifted on a motion stage, and imaged again to simulate a two-source sample with varying separation distance. The resulting measurements were summed digitally to represent a two-source scene. Figure \ref{fig:two-point} shows that fluorescent beads spaced 0.12 super-pixels (20 $\mu$m) apart can be resolved. Notably, the experimental resolution exceeds the theoretical prediction due to the sparsity prior used in reconstruction.
    
\begin{figure}[htb!]
    \centering
    \includegraphics[width=1\linewidth]{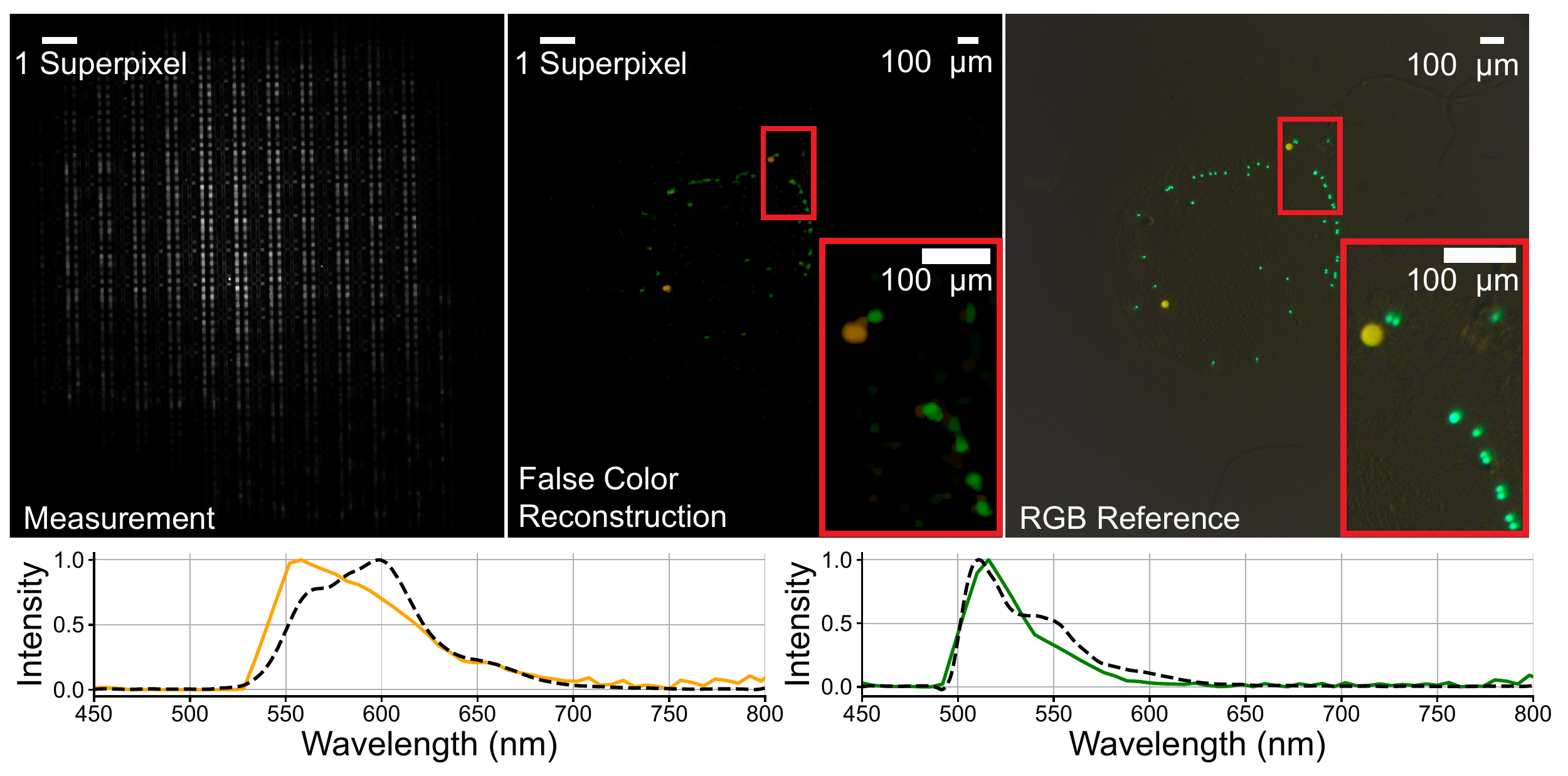}
    \caption{A sample with 10 µm green and 30 µm orange fluorescent beads imaged with a 4$\times$ objective demonstrates resolved features within a single super-pixel. Left: measurement (contrast-stretched for visualization). Center: false-color reconstruction with zoomed inset. Bottom: learned fluorophore spectra compared against ground truth. Right: spatial ground truth captured on a color camera.}
    \label{fig:multipoint}
\end{figure}

Resolution in computational imaging systems is also scene-dependent.  To test performance on a complex scene, we imaged a sample of 10 µm green and 30 µm orange fluorescent beads. The measurement and reconstruction are shown in Fig.~\ref{fig:multipoint}. Beads immediately next to each other located closer than the theoretical resolution limit could not be distinguished, but the reconstructed spectra closely matched the ground truth. 

To evaluate system performance with a denser, more complex sample, we imaged the fluorescent digits on a negative US Air Force resolution target (Edmund Optics).

Note that this system is not diffraction-limited by the collection objective, and will have a lower spatial resolution than a traditional single channel fluorescent microscope with the same objective (ex: ~30 micron vs 3 micron for a 4$\times$ objective).  

\subsection{Dynamic sample}
\begin{figure}[tbh]
    \centering
    \includegraphics[width=1\linewidth]{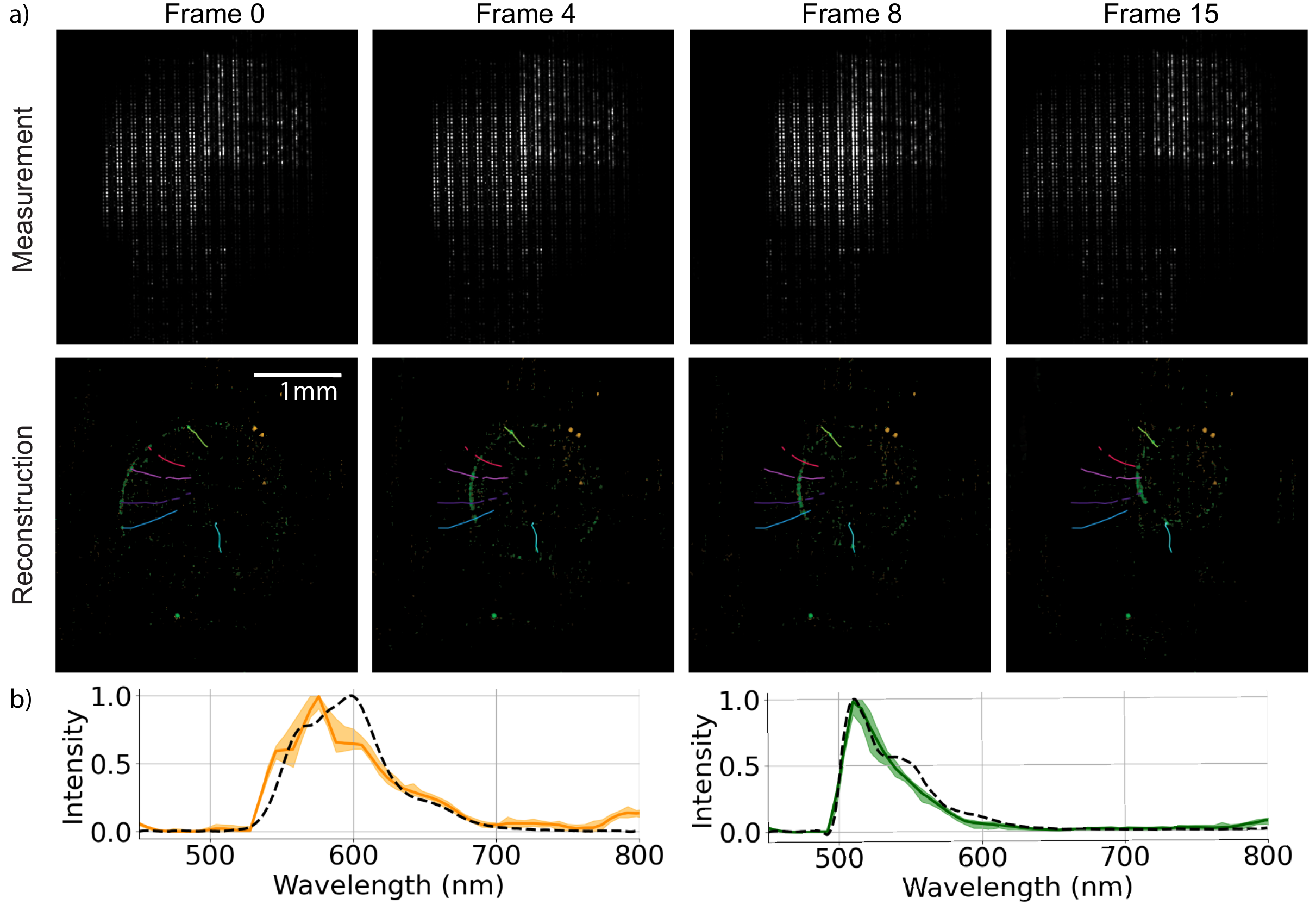}
    \caption{Several time frames of a reconstructed hyperspectral video of a drying water droplet with fluorescent beads in it. a) Top row shows the measurements and below it is the spatial reconstruction of the corresponding frame.  Particle tracking was done on the reconstructed frames to show the movement of the beads.  b) The reconstructed spectra of the orange and green beeds in the scene. The spectra of a single bead over time was captured and remains consistent. }
    \label{fig:video}
\end{figure}

A primary advantage of a snapshot system is the ability to capture videos of dynamic samples. To demonstrate this, a drying water droplet containing fluorescent beads was imaged. As the droplet dried, the beads shifted positions, moving in and out of the focal plane. Sequential frames were acquired over a 10-minute period with 2 second exposure per frame using a 4$\times$ objective. The hyperspectral reconstructions of selected frames are shown in Fig.~\ref{fig:video}. Particle tracking was performed on the reconstructed frames. As the droplet dried, the beads moved closer together, increasing the difficulty of resolving and tracking them individually.

\subsection{MRBLES beads}
\begin{figure}[tbh]
    \centering
    \includegraphics[width=1\linewidth]{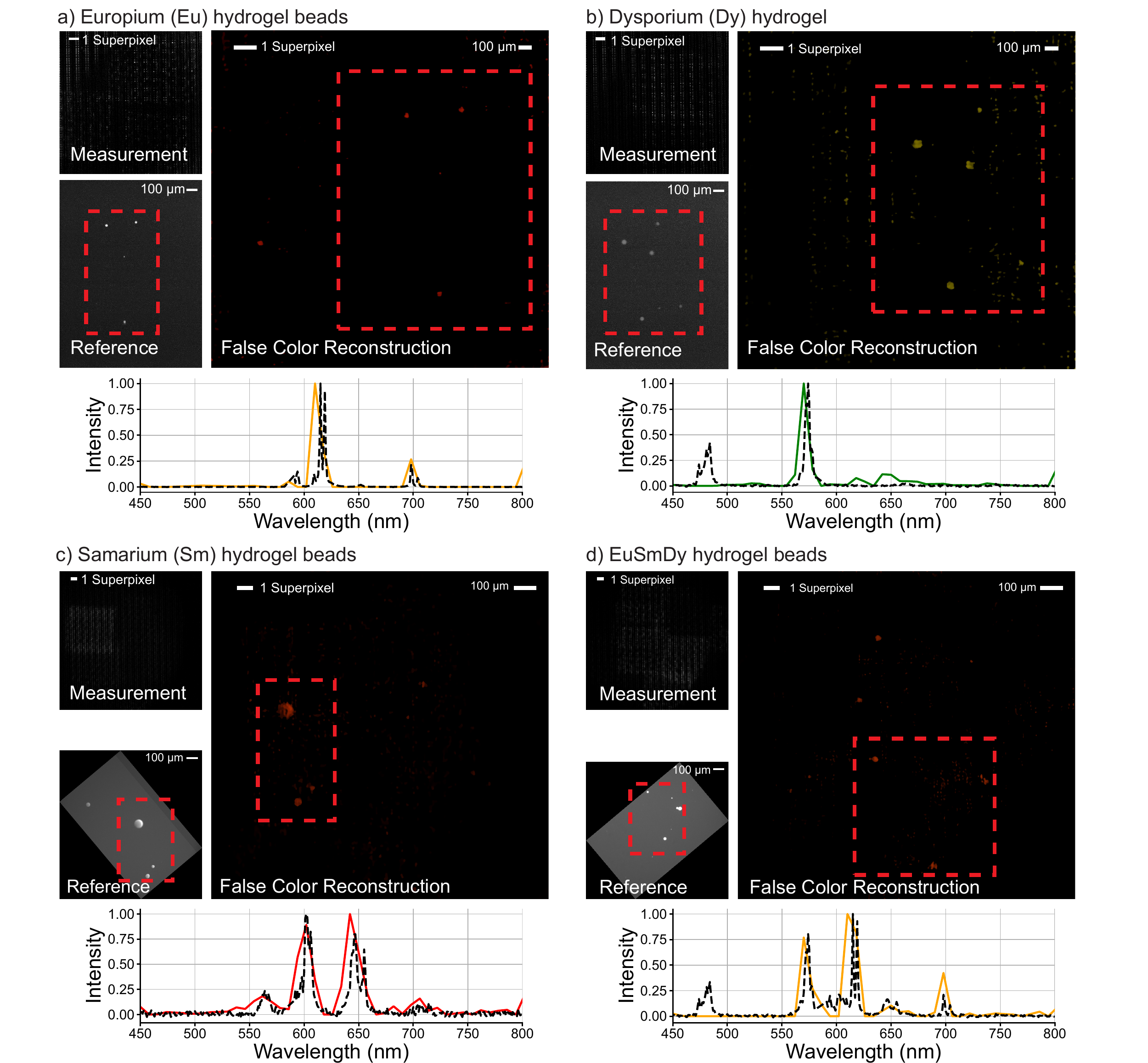}
    \caption{Multiplexed MRBLES hydrogel bioassay beads imaged using our hyperspectral microscope. (a) Europium, (b) Dysprosium, (c) Samarium, and (d) a ratio combination of 10\% Eu, 34\% Sm, and 4.25\% Dy. The measurement is shown on the left, with the false-color reconstruction on the right. The spectra of the beads are compared against the ground truth on the bottom. The beads range from 20–50 µm in diameter. Measurements were acquired with a 60-second exposure using a 4$\times$ (a and b) or 10$\times$ objective (c and d).  A long-pass filter in the emission path rejects any peaks below below 500 nm from reaching the sensor, hence are not reconstructed. }
    \label{fig:mrbles}
\end{figure}

To evaluate performance on a sample with sharper spectral features, MRBLES hydrogel beads were imaged. These beads, used in multiplexed bioassays, represent a set of six distinct spectral codes and vary in diameter (~20–50 µm). Hydrogel lanthanide beads were synthesized according to Reference \cite{Feng2020-rx} with higher concentration of lanthanide dye to increase brightness.  For imaging, the beads were placed on a quartz slide and excited with an ultraviolet (285 nm) LED in trans illumination.  An additional 400 nm long pass filter was placed over the objective lens to to reject UV excitation and limit autofluorescence of the objective's glass.  Ground truth spectra for lanthanides was acquired using a fluorometer (Fluorolog-3). The measurement and reconstruction are shown in Fig.~\ref{fig:mrbles}. The reconstructed bead spectra match the ground truth well within the spectral resolution of the system. Due to the sample's low brightness, an acquisition time of 60 seconds was required.

\subsection{Cells}
\begin{figure}[tbh]
    \centering
    \includegraphics[width=1\linewidth]{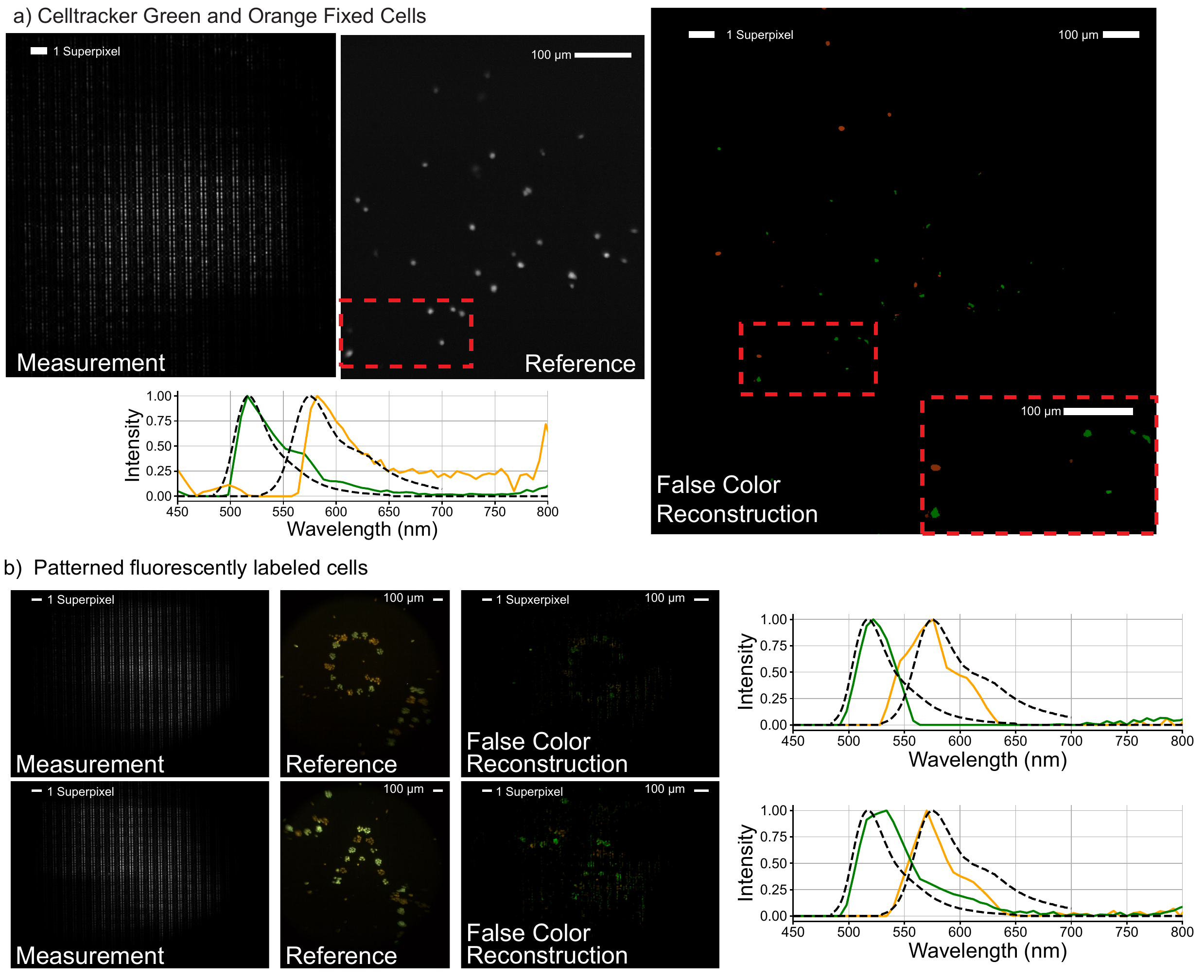}
    \caption{Fixed fluorescently-labeled cells imaged using our hyperspectral microscope, enabling full spectral capture and unmixing. (a) human umbilical vein endothelial cells (HUVEC) labeled with CellTracker Green dye. (b)  HUVEC cells patterned into a CAL shape labeled with Celltracker Green and Celltracker Orange dye.  Multiple fields of view (2 total) were stitched together to capture the whole shape.}
    \label{fig:fixed-cells}
\end{figure}

Fixed fluorescently-labeled human umbilical vein endothelial cells (HUVECs) were imaged to further evaluate system performance when imaging samples with overlapping emission spectra. The fixed cell samples were cultured and fluorescently labeled using CellTracker Green and CellTracker Orange (ThermoFisher). Ground truth spectra for the fluorescent dyes were obtained from SpectraViewer on ThermoFisher's website. The measurement and reconstruction are shown in Fig.~\ref{fig:fixed-cells}. The reconstructed spectra were compared with the vendor-provided ground truth spectra, demonstrating consistency within the spectral resolution of the system.  The CellTracker Orange dye was ~10$\times$ dimmer than the CellTracker Green dye.  Consequently, we choose to take advantage of our seperable recon and increase the intensity of the orange channel by 10$\times$ to improve visibility.

\subsection{Dense samples}
Lastly, we tried to image more complex scenes.  We imaged patterned HUVEC cells in the shape of "CAL".  These cells were labeled with CellTracker Green and CellTracker Orange. Single-cell patterning was achieved with a photolithography-based process called high-throughput DNA-directed patterning (htDNA-dp)~\cite{Scheideler2020-sn} to create the "CAL"-shaped cell sample. HtDNA-dp employs photolithography to selectively expose regions of an aldehyde-coated glass slide, which are subsequently incubated with single-stranded DNA oligonucleotides. Through reductive amination, the oligonucleotides bind to the exposed areas. Cells tagged with complementary single-stranded DNA oligos are then flowed over the slide, where they hybridize to the patterned regions. This technique enables precise, high-resolution spatial positioning of individual cells with high throughput. For this study, three distinct photolithography masks and oligo sequences were used to pattern the letters "CAL." 

The reconstructed scene and spectra are shown in Fig.~\ref{fig:fixed-cells}b.  The reconstructed spectra matches the vendor-provided ground truth.  The spatial reconstruction has artifacts due to the low contrast in the sample between the HUVEC cells and the background. 

In addition to being significantly dimmer, these denser fluorescent samples activate fewer spectral channels compared to previous broadband scenes imaged by Monakhova \textit{et al} in the photography application \cite{Monakhova2020-gd}.  Thus, there are not as many activated sensor pixels making spatial recovery more difficult.  


\section{Discussion}
The Spectral DiffuserScope has been demonstrated as an effective method for simultaneously capturing spatial and spectral information about dynamic fluorescence microscopy samples. A primary advantage of this approach is its compact and simple design, which integrates easily with existing microscopes. The attachment can be built with minimal effort, enabling hyperspectral sample collection. Additionally, the hyperspectral imager attachment is portable and can function as a standalone imaging system when magnification is not required. Demonstrated applications include particle tracking, bead imaging, and cell imaging. The system is well-suited for use in microfluidics and other lab-on-a-chip applications. Beyond biological samples, it is also applicable to materials science and other fields, provided the samples are sparse and sufficiently bright.

 The absence of moving parts makes our system simpler than a filter wheel-based fluorescence microscope, which remains the most common approach today. Additionally, we provide significant acquisition speed advantages: while filter wheel systems require N separate exposures (one per wavelength), each with its own exposure time plus mechanical switching overhead, we capture all spectral information simultaneously in a single exposure. However, this speed advantage comes with a tradeoff in photon efficiency. In our snapshot system, light from each wavelength is captured only a subset of the sensor pixels due to the filter array, whereas filter wheel systems dedicate the full sensor area to each wavelength sequentially. Consequently, for equivalent signal-to-noise ratios per spectral channel, longer exposures may be required in our snapshot system to compensate for the reduced photon collection efficiency per wavelength. Our approach remains advantageous for dynamic samples and high-throughput applications where acquisition speed outweighs SNR considerations as well as bright samples.

The camera's sensitivity posed a significant challenge in our experiments. The system used a cost-effective, room-temperature CMOS camera modified with a spectral filter array attached. Filter wheel-based fluorescence systems typically allow individual adjustment of exposure time for each channel and use higher grade scientific cooled CMOS sensors. In our system, weaker channels shared the same exposure as stronger, making it more difficult to visualize weaker signals without saturating the stronger ones. A camera sensor with higher dynamic range or pixel-adjustable exposure settings would address this limitation. In particular, a cooled scientific CMOS (sCMOS) camera could improve noise performance and sensitivity. However, sCMOS sensors are more expensive and it is generally not desirable or straightforward to remove the coverglass on the sensor. While the design can work with the cover glass intact, it tends to introduce angle sensitivity in the spectral calibration and alters the forward model, which should be accounted for in the inverse problem~\cite{Raniwala2023-zt, Raniwala2023-pc}.

Stray light was another major limitation, causing reconstruction artifacts. Measurements were conducted using an older TE300 microscope, which likely had damaged fluorescence filters contributing to stray light. Additionally, reflections between the spectral filter glass and the diffuser introduced reconstruction artifacts. Anti-reflective coating on the back of the spectral filter array's glass substrate could mitigate these issues in the future. Optical alignment also played a critical role in maintaining spatial invariance across the field-of-view. Spatial invariance was verified visually by inspecting the point spread function across the sensor's unfiltered side. The spectral filter was not pixel-aligned to the camera, which necessitated experimental calibration with a monochromator to get the spectral response of each individual camera pixel. A pixel-aligned filter in future designs would reduce calibration requirements and improve overall system performance.

Reconstruction resolution and quality is scene complexity dependent. The system performed best with bright, sparse objects, while dense narrowband fluorescent scenes failed to reconstruct because the compression ratio was too high. Dense scenes required broader spectral features to activate more pixels. For example,  we tried imaging the bars of a green fluorescent USAF resolution target (\ref{sec:supp_results:usaf}) and were unsuccessful because the emission was too narrowband to activate enough pixels. Choosing a spectral filter array tailored to the fluorophores being imaged would significantly enhance performance. A custom-designed spectral filter array optimized for the specific emission spectra of the fluorophores would improve light throughput and sensitivity, thereby reducing acquisition times, particularly for dim samples such as MRBLES beads.

\textbf{Future work}
To assist with low-light imaging, this work utilized a custom-designed diffuser. Future developments could jointly optimize the diffuser and spectral filter array through data-driven, end-to-end design, further enhancing light throughput and system performance.

The droplet samples were inherently 3D, but the reconstruction assumed a 2D model, leading model mismatch as beads moved in and out of the focal plane. In the future, a 3D hyperspectral reconstruction algorithm could be implemented. This would require acquiring a PSF stack at different focal planes and modifying the forward model to incorporate a 4D datacube, ($x$, $y$, $z$, $\lambda$), leading to significantly increased computational requirements.

Additionally, parameterized representations such as coordinate-based multi-layer perceptron (MLP) networks or Gaussian splatting could be employed to reduce the number of parameters in the reconstruction. These approaches could also impose stronger priors or enforce continuity between frames of dynamic samples, potentially improving reconstruction fidelity and particle tracking accuracy. In this work, a physics-based classical reconstruction algorithm with sparsity constraints was deliberately chosen over a deep learning-based approach to maintain the system's agnosticism to the types of samples it may encounter. Collecting a diverse dataset of fluorescent hyperspectral microscopy samples would assist with the training of data-driven approaches.

We provided the bill of materials and assembly instructions for the hyperspectral camera imager attachment and the open soure code for the reconstruction pipeline in the Supplement.  We also mentioned several improvements to the hardware and reconstruction pipeline to push performance further.  We hope these efforts will enable others to build on this work. 

Overall, the Spectral DiffuserScope presents a promising approach for hyperspectral microscopy, with potential for broader use cases and further optimization.

\begin{backmatter}

\bmsection{Author credit statement}

Neerja Aggarwal and Eric Markley (equal contribution) led the project conceptualization, optics and hardware setup, reconstruction algorithm and code, experimental capture and reconstruction, and the manuscript writing and editing. Kyung Chul Lee designed the custom phase mask.  Junghyun Bae, Nakkyu Baek fabricated the custom phase mask.  Wook Park and Seung Ah Lee provided funding and supervision for the custom phase mask.  William Houck helped provide the spectral filter array.    Minsung Cho creating the MRBLES samples.  Youngbim Lim assisted with imaging MRBLES samples.  Polly Fordyce provided funding and supervision for MRBLES samples.  Stephanie Eberly created the HUVEC cell samples.  Lydia Sohn provided funding and supervision for the patterned HUVEC cell samples.  Kristina Monakhova assisted with project conceptualization and manuscript review.  Laura Waller provided project conceptualization, supervision, funding acquisition, and manuscript review and editing.
 
\bmsection{Acknowledgments}
We thank Kyrollos Yanny for assistance with debugging reconstructions.  We thank Sean Kitayama for assistance designing the patterned cell samples.  We thank Yashovardhan Raniwala for assistance with exploring alternative optical setups.  We thank Christian Foley for assistance with spectral calibration.  We thank Caroline Horn for collecting ground truth lanthanide spectra.  We thank  Viavi Solutions for providing the spectral filter arrays for this project. 

\bmsection{Disclosures}
William Houck is an employee at Viavi Solutions, manufacturer of spectral filter arrays. The remaining authors declare no conflicts of interest.

\bmsection{Data availability} Data underlying the results presented in this paper will be made available upon request.  The code for this project is open source at \url{https://github.com/neerja/spectraldiffuserscope}

\bmsection{Supplemental document}
See Supplementary Information, included at the end of this document, for supporting content.

\end{backmatter}


\bibliography{erics_zotero, paperpile}

\clearpage
\nolinenumbers
\setcounter{section}{0}
\setcounter{table}{0}
\setcounter{figure}{0}
\makeatletter
\renewcommand\thesection{S\@arabic\c@section}
\renewcommand\thesubsection{\thesection.\@arabic\c@subsection}
\renewcommand\thesubsubsection{\thesubsection.\@arabic\c@subsubsection}
\renewcommand\thetable{S\@arabic\c@table}
\renewcommand\thefigure{S\@arabic\c@figure}
\makeatother

\begin{center}
{\large\bfseries Supplementary Information for Spectral DiffuserScope}
\end{center}
\bigskip

\section{Supplementary methods}\label{sec:supp_methods}

\subsection{Bill of materials}\label{sec:supp_methods:hardware_bom}

\begin{table}[h!]
    \centering
    \begin{tabular}{|l|l|c|c|}
         \hline
            \textbf{Item} & \textbf{Part Number} & \textbf{Quantity} &\textbf{Identifier} \\
            \hline
            Zeiss microscope sideport c-mount adapter & Zeiss 60N-C & 1 & A \\
            Cmount internal to SM1 external adapter & Thorlabs SM1A10 & 1 & B \\
            30mm cage plate  & Thorlabs CP33T & 1 & C \\
            500 nm long pass filter & Thorlabs FELH0500 & 1 & D \\
            XY translation cage plate & Thorlabs CXY1A & 1 & E \\
            SM1 external to M25x0.75 internal adapter & SM1A12 & 1 & F \\
            Achromatic scan lens & Thorlabs LSM03-VIS  & 1 & G\\
            30mm to 60mm cage plate adapter & Thorlabs LCP33 & 1 & H\\
            SM1 0.5" lens tube & Thorlabs SM1L05 & 1 & I \\
            2.5 mm pinhole & Thorlabs P2500K & 1 & J \\
            60mm translating cage segment plate & Thorlabs LCPX1 & 1 & K \\
            3" cage rods & Thorlabs ER3 & 4 & L \\
            1" cage rods & Thorlabs ER1 & 8 & M \\
            Optical post & Thorlabs TR1.5 & 1 & N \\
            Optical post holder & Thorlabs UPH1.5 & 1 & O \\
            Camera adapter plate & see link below & 1 & P \\
            Diffuser & see text below & 1  & Q \\
            Spectral filter array & Viavi Solutions  & 1 & R \\
            Board level image sensor & TIS DMM 37UX178-ML & 1 & S \\
            \hline
    \end{tabular}
    \caption{Parts list for Spectral DiffuserScope setup}
    \label{tab:parts_list}
\end{table}

\begin{figure}[h!]
    \centering
    \includegraphics[width=0.95\linewidth]{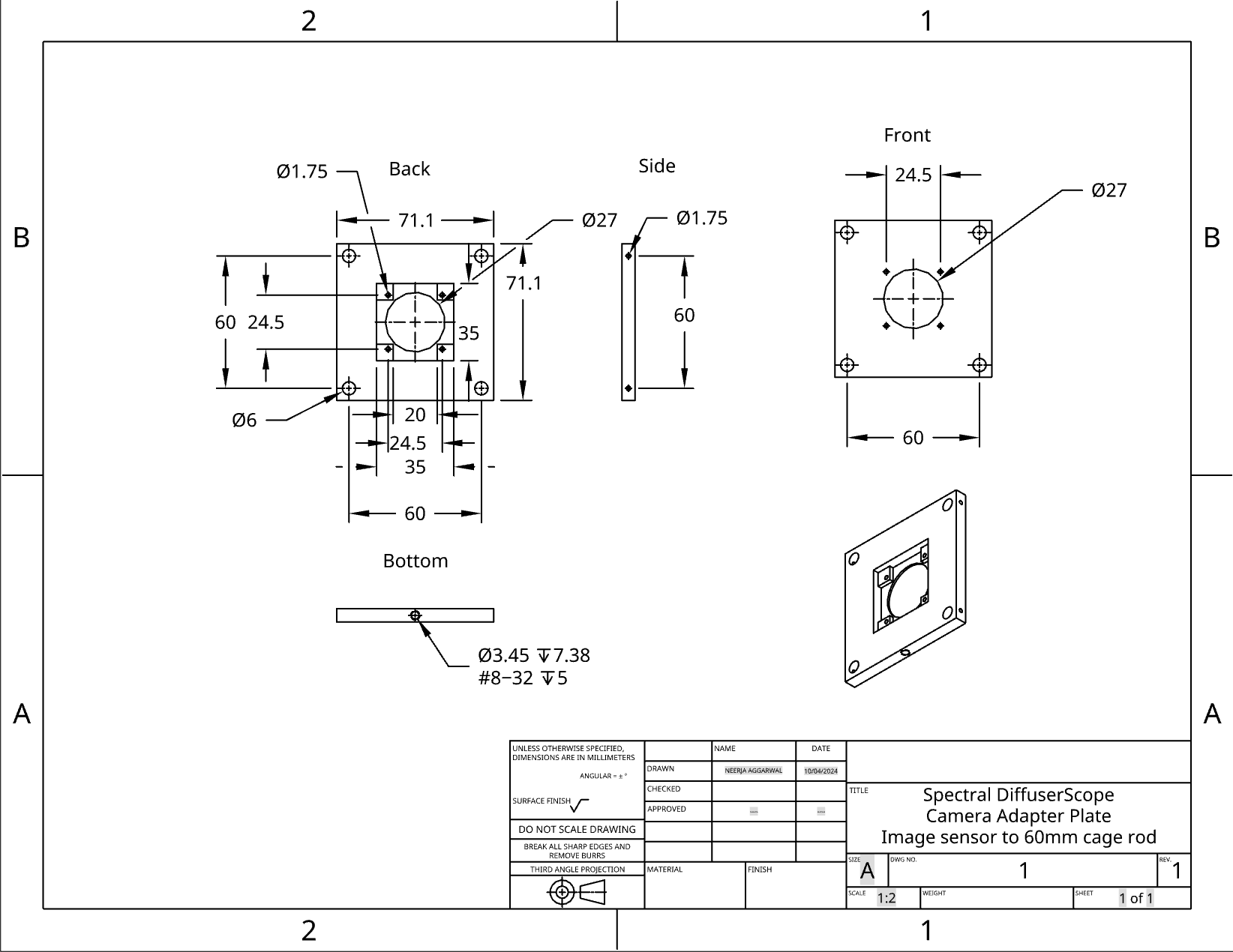}
    \caption{Drawing for the camera adapter plate.  All dimensions in millimeters.}
    \label{fig:camera_adapter_plate}
\end{figure}

CAD file for camera adapter plate is uploaded to the data directory of this project and also available here: \url{https://cad.onshape.com/documents/449caf9366b28bb97cd6dcee/w/f675872992f46a11f2c93e37/e/8b24da2db02d465414c0caa0}  The plate was CNC machined from aluminum through a rapid prototype service.   See Figure \ref{fig:camera_adapter_plate} for a drawing of the camera adapter plate.

We used a custom engineered diffuser from our collaborators. Alternatively, an off the shelf Luminit 0.5 degree holographic diffuser can be obtained from EdmundOptics.

\subsection{Usage Instructions}
\begin{enumerate}
    \item Assemble the spectral camera according to the instructions in main text Section \ref{sec:methods:spectral_camera_assembly}.
    \item Calibration the spectral filter matrix according to instructions in main text Section \ref{sec:methods:calibration} and Supplementary Info \ref{sec:supp_methods:spectral_calibration}.
    \item Assemble using the instructions in Supplementary Info \ref{sec:supp_methods:hardware_assembly}.
    \item Calibrate the diffuser point spread function using the instructions in main text Section \ref{sec:methods:calibration}.
    \item Slide the spectral camera using the 60mm translation plate (part K) so that the point spread function now lands on the center of the spectral filter array.
    \item The setup is now ready for imaging.  Place a fluorescent sample in the microscope and acquire image using TIS software, Micromanager, or other software.  Make sure to also collect a background image ideally from an empty part of the sample slide for subtraction.
    \item Preprocess the calibration data using the instructions in main text  \ref{sec:methods:calibration} and in the spectral calibration notebooks in the Github repo referenced.
    \item Run the hyperspectral datacube reconstruction code in the Github repo referenced.
\end{enumerate}

\subsubsection{Optical setup assembly instructions}\label{sec:supp_methods:hardware_assembly}
\begin{figure}[h!]
    \centering
    \includegraphics[width=0.8\linewidth]{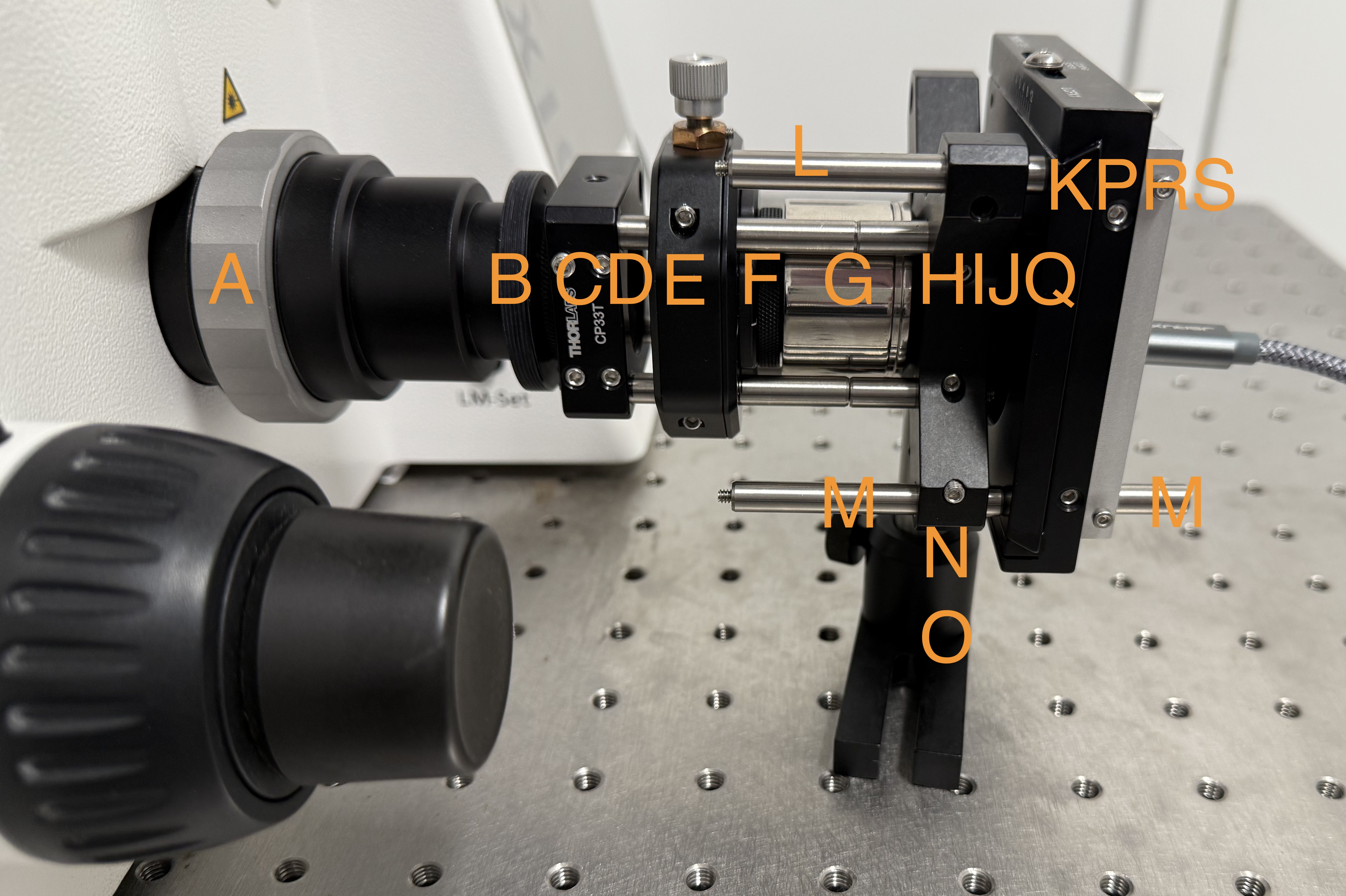}
    \caption{Annotated image of the optical setup with labeled parts.  Note, there are variations of parts B and M in the image.}
    \label{fig:setup_annotated}
\end{figure}

\begin{enumerate}
    \item Assemble the spectral camera according to the instructions in main text Section \ref{sec:methods:spectral_camera_assembly}.
    \item Calibration the spectral filter matrix according to instructions in main text Section \ref{sec:methods:calibration} and Supplementary Info \ref{sec:supp_methods:spectral_calibration}.
    \item Attach the Zeiss microscope sideport c-mount adapter (part A) to the microscope sideport.
    \item Attach the C-mount internal to SM1 external adapter (part B) to part A.
    \item Insert the 500 nm long pass filter (part D) into the 30mm cage plate (part C) and attach to part B.
    \item Attach the achromatic scan lens (part G) to the XY translation cage plate (part E) using the M25x0.75 to SM1 adapter (part F).
    \item Attach part EFG to part C using cage rods (part L).
    \item Align along x and y.  The microscope adapter and cage system should already be aligned to the optical axis. But if not, align the scan lens to the center of the optical axis by adjusting the XY translation cage plate. This can be done by imaging a bright object in the microscope and select sideport for output.  Remove the scan lens.  Focus the object 1 meter away on a white paper. Place the scan lens back in the setup and adjust the XY translation cage plate until the image is in the same place as before.
    \item Align the scan lens along z (optical axis) so it is one focal length away from the original sideport imaging plane.  This can be done by imaging a bright object.  Focus on the object using the eyepiece and then switch to the sideport.  Place the scan lens roughly 2 focal lengths away.  This should make a 4f imaging system.  Place a monohcrome camera temporarily in the secondary image plane.  Then as you move the scan lens closer to the sideport, the secondary image plane should move farther away.  Adjust the monochrome camera placement to stay in focus.  Once the monochrome camera is 1 meter away, the scan lens is effectively imaging at infinity and is roughly one focal length away from the sideport image plane.
    \item If using a custom engineer diffuser (part Q), you can use a piece of tape with a square aperture to attach the diffuser to the mounted pinhole (part J) and then insert into the lens tube (part I)  If using an off the shelf mounted diffuser, insert directly into part I.
    \item Attach part I to the 30mm to 60mm cage plate adapter (part H).
    \item Attach a post (part N) and post holder (part O) to the bottom of part H. Keep the post holder screw loose until the next step.
    \item Attach the part assembly HIJQ to part E using the previously attached cage rods (part L).  It's important that the cage rods don't stick out too far past part H (hence best to use 3" rods).  Tighten the post holder screw on part O and fasten part O to the optical table.
    \item Attach the board level image sensor (part S) with the glued spectral filter to the camera adapter plate (part P) using 1" cage rods (part M).
    \item Use four of the 1" cage rods (part M) to attach the camera adapter plate (part P) to the 60mm translating cage segment plate (part K).
    \item Use the remaining four 1" cage rods (part M) to attach part assembly KPS to part H.
    \item Align the camera along the optical axis to the focal plane of the diffuser.  This can be done by imaging a 5 micron fluorescent bead on the side of the image sensor unoccluded by the spectral filter and moving the camera until diffuser's caustic ridges are in focus.  We used The Imaging Source's IC Capture software to view the camera feed.
\end{enumerate}

\subsection{Spectral calibration processing}\label{sec:supp_methods:spectral_calibration}
\begin{figure}[H]
    \centering
    \includegraphics[width=0.5\linewidth]{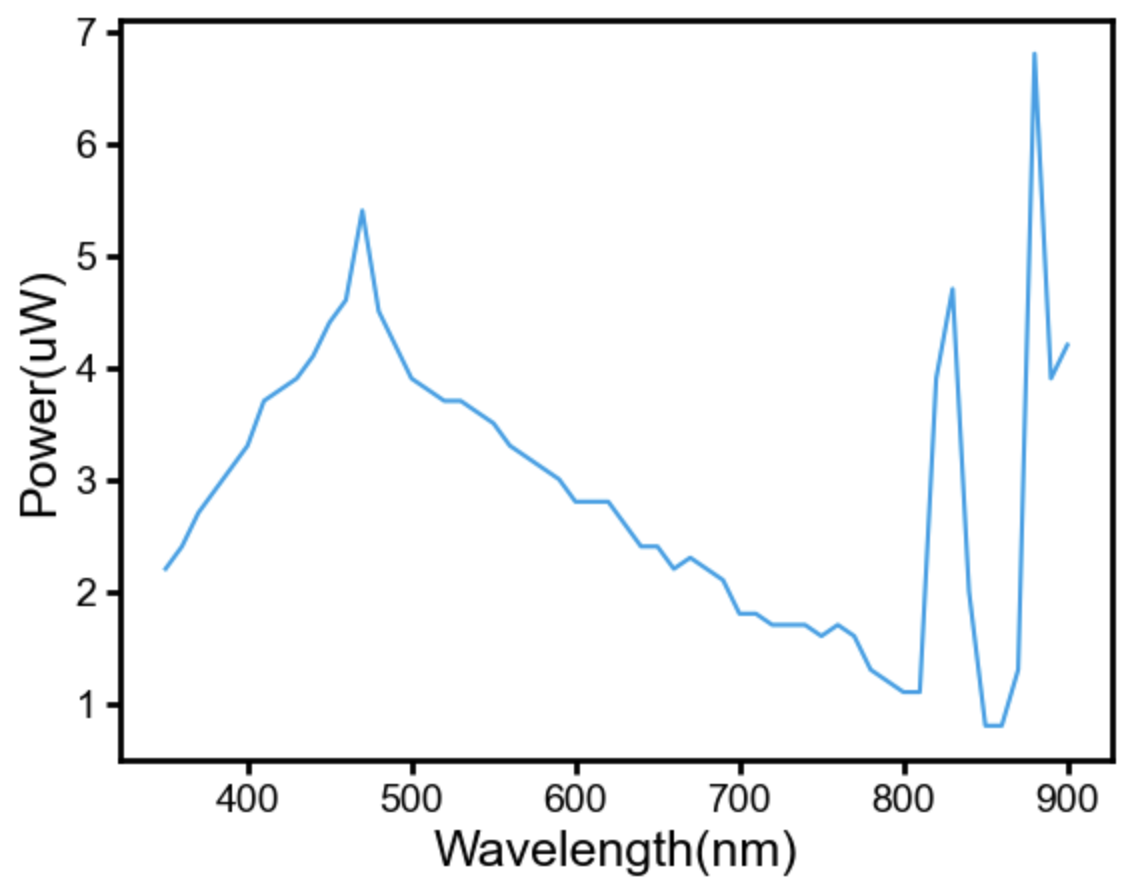}
    \caption{Output power for Cornerstone 130 monochromator}
\end{figure}

\subsection{Diffuser design and optimization}\label{sec:supp_methods:diffuser}

\begin{figure}[H]
    \centering
    \includegraphics[width=1\linewidth]{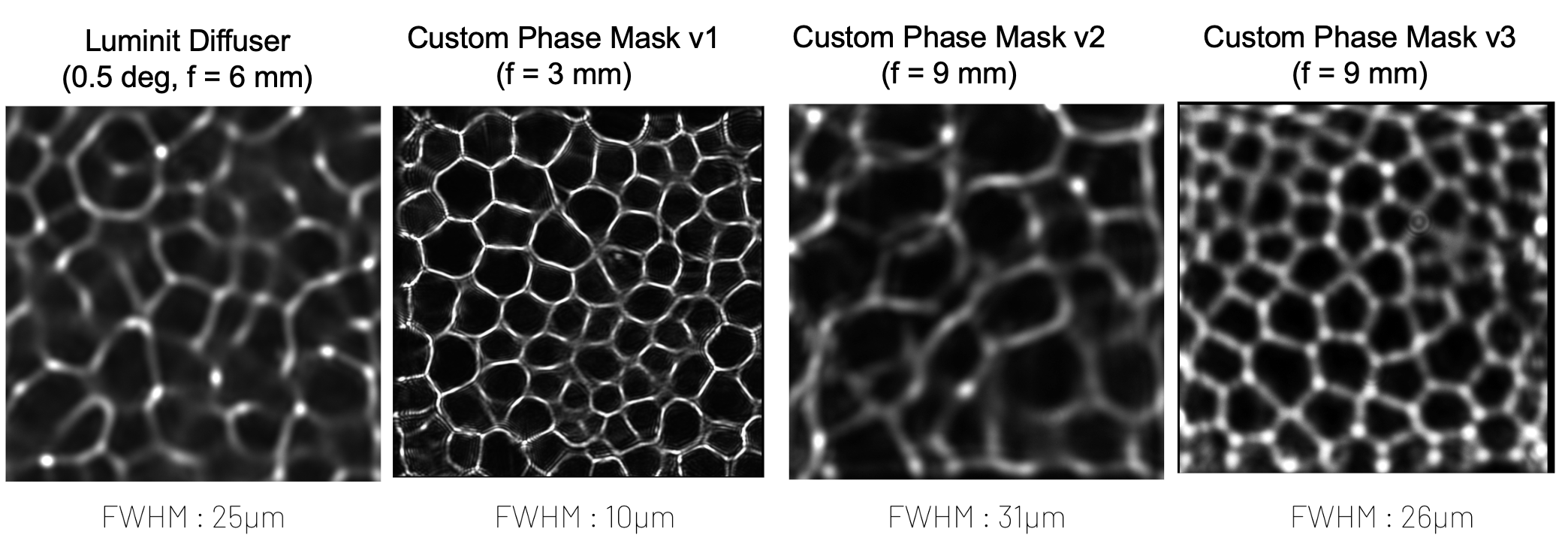}
    \caption{Diffuser point spread functions for various designs. Note, these images were collected according to Reference \cite{Lee2023-os}}
\end{figure}

\section*{Supplementary results}\label{sec:supp_results}


\subsection{USAF  target}\label{sec:supp_results:usaf}
We imaged a digit on the fluorescent USAF resolution target as shown below.  The measurement only had the green filter channels lit up since the sample was narrowband.  This reconstruction relied heavily on a total variation/sparse gradient prior to obtain the sample's dense spatial image.
\begin{figure}[H]
    \centering
    \includegraphics[width=1\linewidth]{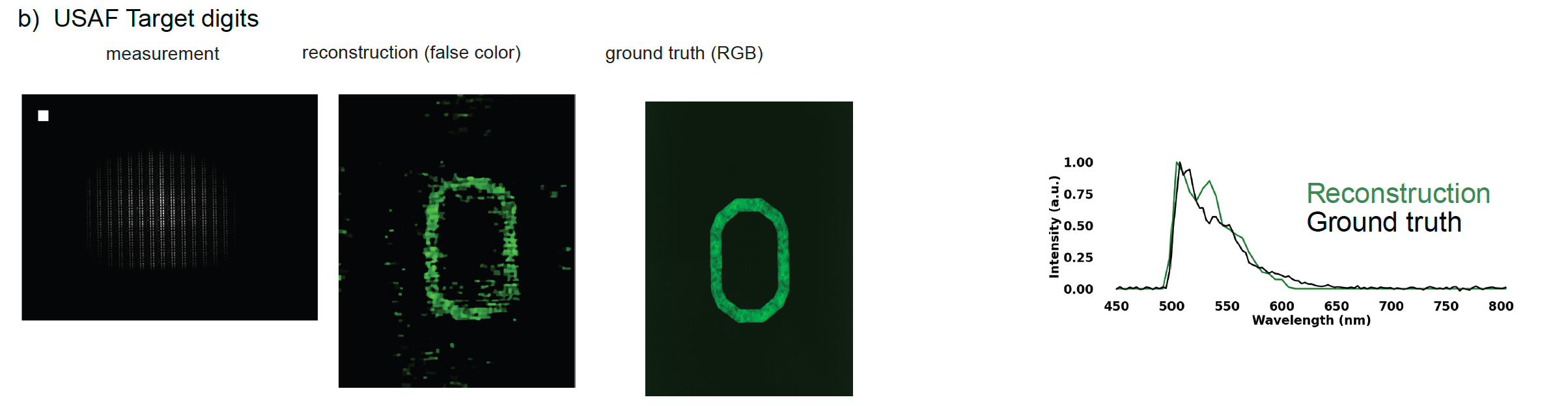}
    \caption{Digit on fluorescent USAF resolution target imaged using hyperspectral microscope}
\end{figure}

\begin{table}[htb]
\centering
\caption{Hyperspectral Imaging Approaches}
\label{tab:hyperspectral}
\resizebox{\textwidth}{!}{%
{\small
\begin{tabular}{>{\raggedright\arraybackslash}p{1.5cm} >{\raggedright\arraybackslash}p{2cm} >{\raggedright\arraybackslash}p{2cm} >{\raggedright\arraybackslash}p{2cm} >{\raggedright\arraybackslash}p{3cm} >{\raggedright\arraybackslash}p{3.5cm}}
\hline
\textbf{Study} & \textbf{Approach} & \textbf{Samples Shown} & \textbf{Channels, Bandwidth} & \textbf{Advantages} & \textbf{Disadvantages} \\
\hline
This work & Multiplexing using phase mask onto spectral filter array + compressed sensing & Fluorescent beads, fixed cells & 64 channels, 350-900nm & Compact, simple setup, filters customizable,  large FOV, can control spatial \& spectral resolution separately, can extend to 3D & Restricted to sparse, bright samples, SNR tradeoff with the number of channels, reconstruction relies on spectral priors \\
\hline
Wu, \textit{Sci Reports}, 2016~\cite{Wu2016-if} & Lightfield camera array with filters & Algae, lymph node, larvae & 25 channels; range depends on filters & Volumetric, live imaging, customizable filters & Large, expensive, complex setup, SNR and size tradeoff with the number of channels \\
\hline
Hedde, \textit{Nat Comms Bio}, 2021~\cite{Hedde2021-iz} & Light sheet + filters & Fixed zebrafish eye, live cells, mouse tissue metabolites & 2 channels, 400--700 nm & Only two filters allow high sensitivity, simple setup & Need a full spectral basis of the sample * Not actually a hyperspectral method \\
\hline
Cull, \textit{App Optics}, 2010~\cite{Cull2010-ky} & Coded aperture dispersion + compressed sensing & Fluorescent beads & 22 channels, 490--700 nm & Adaptable to benchtop setups & Limited channels - Reconstruction relies on spectral priors \\
\hline
Lavagnino, \textit{Biophys J}, 2016~\cite{Lavagnino2016-us} & Image dispersion onto a large sensor & Live pancreatic islets & 60 channels & Live samples, subsecond temporal resolution & Limited magnification, long assembly \\
\hline
Orth, \textit{Optica}, 2015~\cite{Orth2014-fx} & Microlens array and prism to disperse onto a large sensor & Full well HeLa cells and fluorescent beads & 13 channels *adjustable & Large FOV, targeted for drug discovery & Limited spectral channels \\
\hline
Zhang, \textit{Nat Methods}, 2015~\cite{Zhang2015-lp} & Prism to disperse along one direction + large sensor & Organelles inside fixed cells & 650--800 nm, \textasciitilde 25 channels & Single molecule sensitivity & Limited to sparse samples - Not truly a snapshot since need multiple exposures to build a full image \\
\hline
\end{tabular}
}
}
\end{table}

\end{document}